\documentclass[aps,prl,twocolumn]{revtex4}

\usepackage{graphicx}

\usepackage{epsf}
\usepackage{psfig}
\usepackage{amsmath}

\newcommand{\re}[1]{(\ref{#1})}

\newcommand{\up}{\uparrow}

\newcommand{\dn}{\downarrow}

\newcommand{\beg}{\begin{equation}}
\newcommand{\en}{\end{equation}}

\newcommand{\eps}{\epsilon}
\newcommand{\lam}{\lambda}

\begin{document}

\title{Dynamical vanishing of the order parameter in a fermionic condensate}
\author{Emil A. Yuzbashyan$^{1}$ and Maxim Dzero$^{2}$}
\affiliation{$^1$Center for Materials Theory, Department of Physics and Astronomy,
Rutgers University, Piscataway, New Jersey 08854, USA \\
$^2$Department of Physics and Astronomy, Iowa State University and Ames
Laboratory, Ames, IA 50011}

\begin{abstract}
We analyze the dynamics of a condensate of
ultra-cold atomic fermions following an abrupt change
 of the pairing strength.  At long times,
the system goes to a non-stationary steady state, which we
 determine exactly.
The superfluid order parameter
asymptotes to a constant value. We show that the order parameter {\it vanishes}  when the
 pairing strength  is decreased below a certain critical value.
 In this case, the steady state of the system combines properties of
 normal and superfluid states -- the  gap and
 the condensate fraction vanish, while the superfluid density is nonzero.
\end{abstract}

\maketitle

Recently, several remarkable experiments have demonstrated Cooper pairing in cold atomic
Fermi gases[\citealp{exp1}--\citealp{exp2}].
Key signatures of  a paired state -- condensation of Cooper pairs\cite{exp1,exp4} and the
pairing gap\cite{exp2}   have been observed.
In addition, trapped gases provide a unique tool to explore  aspects of fermion pairing normally  inaccessible in superconductors.  One of the most exciting prospects is a study of far from equilibrium coherent dynamics of  fermionic
condensates[\citealp{barankov}--\citealp{Emil05}], made possible due to the precise experimental control over interactions between atoms
\cite{exp5,ferm1}.  The dynamics can be initiated by quickly changing  the  pairing  strength with external magnetic field.

\begin{figure}[h]
\centerline{\psfig{file=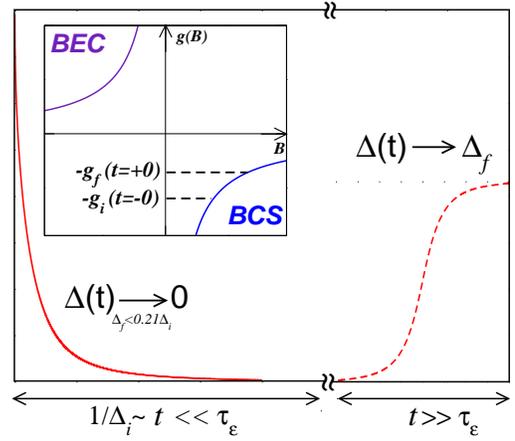,height=6.5cm,width=8.5cm,angle=-90}}
\caption{Time evolution of the BCS order parameter $\Delta(t)$ following an abrupt change of the
coupling constant. The coupling is changed by varying the magnetic field on the BCS side of the
Feshbach resonance (inset).   The order parameter   exponentially decays to zero (solid line) at times
$t\sim 1/\Delta_i$ and the system goes into a {\it gapless} steady state. If the system is continuously cooled, the order
parameter recovers to its equilibrium value $\Delta_f$ (dashed) at times longer than the quasiparticle relaxation time $\tau_\eps$.
 }
\label{Fig1}
\end{figure}

In the present paper, we determine the time evolution of a  fermionic condensate in response to a sudden change of interaction strength. Initially, the gas is in equilibrium at zero temperature on the BCS side of the Feshbach resonance with a coupling constant  $g_i>0$.
At $t=0$ the coupling is
suddenly changed to a smaller value $g_f>0$ on the same
side of the resonance, $g_i\to g_f$, Fig.~\ref{Fig1} (inset). Ground states of the system at the old, $g_i$, and new, $g_f$, values of the coupling
are characterized by corresponding BCS gaps, $\Delta_i$ and $\Delta_f$, respectively. We consider the case $\Delta_i\ge\Delta_f$. It has been shown previously
that
following the change of coupling, the time-dependent order parameter
$\Delta(t)$ asymptotes  to a constant value\cite{Emil05}, $|\Delta(t)|\to\Delta_\infty$ on a timescale
$\tau_\Delta=1/\Delta_i$.
Here we evaluate $\Delta_\infty$
in terms of $\Delta_i$ and $\Delta_f$.

We show that when the coupling is decreased below a certain critical value, $\Delta_\infty$
{\it vanishes}, Fig.~\ref{Fig1}. On a $\tau_\Delta$ timescale  the system goes to a steady {\it non-stationary} state that combines properties of normal and
superfluid states in a peculiar way. For example, the gap vanishes, while the superfluid density remains
finite. Provided  the system is continuously cooled,  the BCS ground state with a gap $\Delta_f$ is reached on the energy relaxation timescale $\tau_\eps$, which is typically much larger than $\tau_\Delta$. Experimental signatures
of the novel  state include the absence of the gap in rf absorbtion spectrum and zero condensate fraction
after a fast projection onto the Bose-Einstein Condensation (BEC) side (see below).

At times $t\ll\tau_\eps$, dynamics of the condensate in the weak coupling regime can be described by the BCS model.
Here we are interested in the thermodynamic limit, in which case one can use
  the BCS mean-field approach\cite{Anderson58}.
Using Anderson's pseudospin representation\cite{Anderson58}, one can describe the mean-field evolution  by a {\it classical} spin Hamiltonian\cite{Anderson58,barankov,YAKZ05}
\beg
H=\sum_j 2\eps_j s_j^z-g\sum_{j,k} s_j^+s_k^-,
\label{bcs}
\en
where $\eps_j$ are single-particle energies relative to the Fermi level and $s_j^\pm=s_j^x\pm is_j^y$. The summation in Eq.~\re{bcs} is over $|\eps_j|<E_F$,
where $E_F$  is the Fermi energy.
Dynamical variables ${\bf s}_j$ are vectors of fixed length, $|{\bf s}_j|=1/2$.
 The BCS order parameter  is $\Delta(t)=\Delta_x-i\Delta_y=g\sum_j s_j^-$.
 Equations of motion  for classical spins ${\bf s}_j$   are
\beg
\dot {\bf s}_j={\bf b}_j \times {\bf s}_j, \phantom{m} {\bf b}_j=\left(-2\Delta_x, -2 \Delta_y, 2\eps_j\right).
\label{spins2}
\en
Components of spins   are related
to Bogoliubov amplitudes $u_j$ and $v_j$
\beg
2s_j^z= |v_j|^2-|u_j|^2,\phantom{l} s_j^-=\bar u_j v_j,
\label{bog}
\en

At $t=0$ the system is in the ground state with gap $\Delta_i$. The ground state is obtained by aligning each spin
${\bf s}_j$ antiparallel to its ``magnetic'' field ${\bf b}_j$ \cite{Anderson58} in order to
minimize the total energy \re{bcs} for coupling constant $g=g_i$
\beg
s_j^x(0)=\frac{\Delta_i}{2\sqrt{\eps_j^2+\Delta_i^2}},~s_j^z(0)=-\frac{\eps_j}{2\sqrt{\eps_j^2+\Delta_i^2}},
\label{initsp}
\en
and $s_j^y(t=0)=0$.

At $t=0$ the coupling is changed, $g_i\to g_f$,   and the initial spin configuration \re{initsp} is no longer an equilibrium for $t>0$. To determine
the time evolution of the system, one has to solve Eqs. \re{spins2} with initial conditions \re{initsp}.

We start with a linear analysis. Solving Eqs.~\re{spins2} (with $g=g_f$) linearized around the spin configuration
\re{initsp}, we obtain up to terms of order $\delta\Delta/\Delta_f$
\beg
\Delta(t)=\Delta_f-8\delta\Delta\int_0^{\infty} d\eps \frac{\cos\omega(\eps)t}{\omega(\eps)
\left[ \pi^2 +h^2(\eps)\right]},
\label{deltaweak}
\en
where $\delta\Delta=\Delta_f-\Delta_i$, $\omega(\eps)=2\sqrt{\eps^2+\Delta_f^2}$, and $h(\eps)=\sinh^{-1} (\eps/\Delta_f)$.
In Eq.~\re{deltaweak}, besides the continuum limit, we took   the weak coupling limit $E_F/\Delta_{i}\to\infty$.

The long time behavior of the order parameter in the linear approximation is obtained from Eq.~\re{deltaweak} by stationary phase method (see also Ref.~\cite{Volkov74}),
\beg
\Delta(t)=\Delta_f-\frac{2\delta\Delta}{\pi^{3/2}\sqrt{\Delta_f t}}\cos\left( 2\Delta_f t+ \frac{\pi}{4}\right),
\label{dasm}
\en
At times $t\gg\tau_\Delta$ the gap approaches a constant value $\Delta_\infty=\Delta_f$ to order $\delta\Delta/\Delta_f$.

Even though the gap is constant, the state of the system is non-stationary\cite{Warner05}. According to Eq.~\re{spins2}, at large
times each spin ${\bf s}_j\equiv{\bf s}(\eps_j)$ precesses in its own constant  field ${\bf b}_j=(-2\Delta_\infty, 0, 2\eps_j)$.
For example, for the $x$-component of spins  we derive from the linearized equations of motion,
\beg
\begin{split}
s_x(\eps)=&\frac{\Delta_f}{2\sqrt{\eps^2+{\Delta_f}^2}}\\&-\frac{\delta\Delta\eps}{(\eps^2+\Delta_f^2) \sqrt{\pi^2+h^2(\eps)}} \cos\left[\omega(\eps) t+\phi(\eps)\right],
\end{split}
\label{sxcont}
\en
However, the gap $\Delta(t)=g\sum_j s_j^x(t)$ contains oscillations with many different frequencies. At large
times they go out of phase and cancel out in the continuum limit.

In the non-linear case it can be shown that the gap decays to a constant $\Delta_\infty$ by a similar mechanism\cite{Emil05}. To determine $\Delta_\infty$, we use the exact solution for the dynamics of the BCS model\cite{YAKZ05}.
Consider the following vector function  of an auxiliary
parameter $u$: ${\bf L}(u)=-\hat {\bf z}/{g}+ \sum_j  {\bf s}_j/{(u-\eps_j)}$,
where $\hat {\bf z}$ is a unit vector along the $z$-axis. Using Eq.~\re{spins2}, we obtain
$d{\bf L}^2(u)/dt=0$, i.e. ${\bf L}^2(u)$ is conserved for any $u$.

\begin{figure}[h]
\centerline{\psfig{file=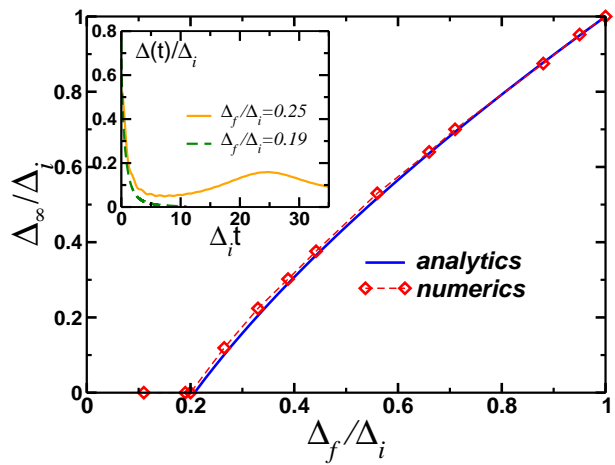,height=6.5cm,width=8.5cm,angle=-90}}
\caption{After a sudden change of the BCS coupling constant, the order parameter
$\Delta(t)$ (inset) saturates  to a constant value
$\Delta_\infty$.
The plot shows the steady state gap $\Delta_\infty$ in units of $\Delta_i$ as a function of the ratio $\Delta_f/\Delta_i$.
The exact result given by Eq.~(\ref{lessinf}) is compared to numerical solution of Eq.~\re{spins2}.
Note that $\Delta_\infty=0$ for $\Delta_f\le 0.2\Delta_i$.
}
\label{d8}
\end{figure}

To determine the steady state gap $\Delta_\infty$, we evaluate ${\bf L}^2(u)$
at $t=0$ for the initial spin configuration \re{initsp} and at $t\gg\tau_\Delta$, when
 each spin ${\bf s}(\eps)$ precesses in  a constant field.
Matching the two expressions, we derive  for $\Delta_f\le\Delta_i$
\beg
\Delta_f= \Delta_i \exp( -\kappa \tan \kappa/2),\qquad \Delta_\infty=\Delta_i\cos\kappa,
\label{lessinf}
\en
where  $0\le \kappa\le \pi/2$.
This equation determines $\Delta_\infty$ in terms of $\Delta_i$ and $\Delta_f$, see Fig.~\ref{d8}.

We make several observations. The   steady state gap   reaches its maximum
$\Delta_\infty=\Delta_f$ at $\Delta_f/\Delta_i=1$. Otherwise, $\Delta_\infty<\Delta_f$. Expanding around
the maximum, we find $\Delta_\infty=\Delta_f-{(\delta\Delta)^2}/{6\Delta_f}$ up to terms of higher order
in $\delta\Delta/\Delta_f$. The linear term vanishes in agreement with Eq.~\re{dasm}.

Most interestingly, we see from Eq.~\re{lessinf} that the order parameter vanishes in the steady state,
$\Delta_\infty=0$, when $\kappa=\pi/2$, i.e. for $(\Delta_f/\Delta_i)_c=e^{-\pi/2}\approx 0.21$.
 If  $\Delta_f/\Delta_i$ is below this critical ratio, Eq.~\re{lessinf} has no solutions -- the system goes into a gapless steady state, $\Delta_\infty=0$,
for $\Delta_f\le 0.21\Delta_i$. We also note from Eq.~\re{lessinf} and Fig.~\ref{d8} that $\Delta_\infty$  has a cusp
at $\Delta_f/\Delta_i=e^{-\pi/2}$, i.e. the transition to the gapless state is second order.

To gain further insight into  properties of the gapless  state,
we need to know the spin configuration in this state.
At large times  spin ${\bf s}(\eps)$ rotates in a constant  field ${\bf b}(\eps)=-2\Delta_\infty\hat{\bf x}+ 2\eps \hat{\bf z}$, where $\hat{\bf x}$ and $\hat{\bf z}$ are unit vectors along the $x$ and $z$ axes, respectively.
 In the gapless case, ${\bf b}(\eps)=2\eps\hat{\bf z}$. The spin configuration can be characterized by the angle $\theta(\eps)$ between ${\bf s}(\eps)$ and $-{\bf b}(\eps)$. This angle can be determined by matching the conserved quantity ${\bf L}^2(u)$ at $t=0$ and $t\gg \tau_\Delta$,
\beg
\sin^2\theta(\eps)=\frac{G(\eps)}{2\pi^2}-\sqrt{ \frac{G^2(\eps)}{4\pi^4}-
\frac{4\beta^2\Delta_i^2}{\pi^2(\eps^2+\Delta_i^2)}}\phantom{a},
\label{sin22}
\en
where $\beta=\ln(\Delta_i/\Delta_f)$ and
\beg
\begin{split}
G(\eps)=&\pi^2+4\left[\sinh^{-1} (\eps/\Delta_i)\right]^2\\&+\frac{8\beta\eps}{\sqrt{\eps^2+\Delta_i^2}}\sinh^{-1} (\eps/\Delta_i) +4\beta^2
\end{split}
\en

\begin{figure}[h]
\centerline{\psfig{file=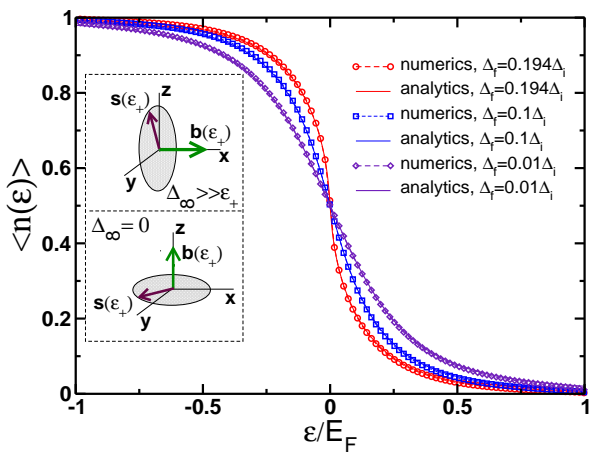,height=6.5cm,width=8.5cm,angle=-90}}
\caption{Plot of the average
occupation number $\langle{n}(\eps)\rangle$  as a function of
energy    in the gapless  state    for various values of $\Delta_f/\Delta_i<e^{-\pi/2}$.
The exact result given by Eq.~(\ref{sin22}) is compared to numerical solution of Eq.~\re{spins2}.
Note that the distribution
function around the Fermi level  is smeared  over a width $\delta\eps\propto\Delta_i$.
The inset shows a spin ${\bf s}(\eps_+)$ at energy $0<\eps_+\ll\Delta_i$ just above and below the critical
point.}
\label{sz}
\end{figure}
Note from Eq.~\re{spins2} that in the steady state the component of spin ${\bf s}(\eps)$ along the field, $s_\parallel(\eps)=-\cos\theta(\eps)/2$, is constant,
while the one perpendicular to the field, $|{\bf s}_\perp|=\sin\theta(\eps)/2$ rotates with a frequency
$\omega(\eps)=2\sqrt{\eps^2+\Delta_\infty^2}$. The approach to the steady state can be studied by linearizing Eq.~\re{spins2} around this state. In particular, we obtain
the asymptotic behavior of $\Delta(t)$
\beg
\frac{\Delta(t)}{\Delta_\infty}=1 + a\frac{\cos(2\Delta_\infty t+\pi/4)}{\sqrt{\Delta_\infty t} },\quad \frac{\Delta_f}{\Delta_i}>e^{-\pi/2}
\label{asymp}
\en
\vspace{-0.5cm}
\beg
\frac{\Delta(t)}{\Delta_i}=A(t) e^{-2\alpha\Delta_i t} +B(t) e^{-2\Delta_i t}, \quad \frac{\Delta_f}{\Delta_i}<e^{-\pi/2}
\label{asymsm}
\en
Here $\alpha=-\cos p$ and $\pi/2\le p \le \pi$ is the solution of $p=\ln(\Delta_i/\Delta_f) \cot(p/2)$. The parameter $\alpha$ has
 a property $\alpha\to0$ when $\Delta_f/\Delta_i\to e^{-\pi/2}$ and
$\alpha\to1$ when $\Delta_f/\Delta_i\to  0$. The coefficient $a$ in Eq.~\re{asymp} is  time-independent, $A(t)$ and $B(t)$ in Eq.~\re{asymsm} are decaying power laws, $A(t), B(t)\propto 1/t^\nu$ with $1/2\le \nu \le 2$.

We conclude that the decay law of $\Delta(t)$ changes from power law to exponential as we cross the critical point.
Above the critical point spins rotate with frequencies $\omega(\eps)=2\sqrt{\eps^2+\Delta_\infty^2}$.
The inverse square root decay and oscillations with frequency $2\Delta_\infty$ in Eq.~\re{asymp} are due to the square root singularity
in the spectral density $d\eps/d\omega\propto \omega/\sqrt{\omega^2-4\Delta_\infty^2}$, cf. Eqs.~(\ref{deltaweak},\ref{dasm}).
Below the critical point $\Delta_\infty$ vanishes, $\omega(\eps)=2\eps$, and the square root anomaly disappears.

What happens to the spin configuration   as $\Delta_f/\Delta_i$ is varied across  the critical
point? Consider a spin at energy $\eps_+\ll\Delta_i$ just above the Fermi energy. Using Eq.~\re{sin22}, we obtain
 $\sin\theta(\eps_+)\approx \sin\theta(0)=2\ln(\Delta_i/\Delta_f)/\pi$ for $\Delta_f>e^{-\pi/2}\Delta_i$ and $\sin\theta(0)=1$ otherwise. The field ${\bf b}(\eps_+)=-2\Delta_\infty\hat{\bf x}+
2\eps_+\hat{\bf z}$  is along the $x$ axis above the critical point ($\Delta_\infty\gg \eps_+$) and along
the $z$ axis below ($\Delta_\infty=0$). At $\Delta_i=\Delta_f$ we have  $\sin\theta(0)=0$, i.e. the spin is parallel to the $x$ axis. As ${\Delta_f}/{\Delta_i}$ decreases,
the $x$ component of the spin also decreases
until it vanishes at the critical point $\Delta_f=e^{-\pi/2}\Delta_i$. Just above it
the  spin lies in the $yz$ plane and rotates
around the  $x$ axis. Below the critical point   it rotates in the $xy$ plane around the
$z$ axis (Fig.~\ref{sz}).

In the gapless steady state each spin rotates around the $z$ axis. Its $z$ component is time-independent,
$s_z(\eps)=-\cos\theta(\eps)/2$. It is related to the average occupation number per fermion species
$\langle \hat n(\eps)\rangle$
at energy $\eps$  as $\langle \hat n(\eps)\rangle=s_z(\eps)+1/2$ \cite{Anderson58}, i.e. $\langle \hat n(\eps)\rangle=(1-\cos\theta(\eps))/2$.
 The distribution
function $\langle \hat n(\eps)\rangle$ is smeared near the Fermi energy over a width $\delta\eps\propto\Delta_i$ (Fig.~\ref{sz}).
Note that in this respect the gapless state is similar to the BCS ground state -- the smearing over a width $\delta\eps\propto\Delta_i$  due to interactions is also
present in the ground state distribution \re{initsp}. In the normal state $s_z(\eps)=-\mbox{sgn } \eps/2$ and $\langle \hat n(\eps)\rangle=\Theta(-\eps)$.

Given   relation \re{bog}, one can reconstruct the time-dependent condensate wave function,
and evaluate normal and anomalous correlation functions, e.g.
\begin{eqnarray}
&
\langle{\hat c}_{\eps\sigma}(t){\hat c}_{\eps\sigma}^\dagger(t')\rangle= e^{i\eps(t'-t)} \cos^2\frac{\theta(\eps)}{2}\label{greenfns1}\\
& \langle{\hat c}^\dagger_{\eps\up}(t){\hat c}^\dagger_{\eps\dn}(t')\rangle=e^{i\eps(t'+t)} \cos\frac{\theta(\eps)}{2}
\sin\frac{\theta(\eps)}{2}
\label{greenfns2}
\end{eqnarray}
where $\hat c_{\eps\sigma}$ ($\hat c_{\eps\sigma}^\dagger$) annihilates (creates) a fermion of
one of the two species $\sigma=\up, \dn$ on energy level $\eps$. Note that even though $\Delta_\infty$ vanishes,  anomalous averages do not.

Next, we determine the superfluid density $n_s$ in the gapless state.
Consider a degenerate Fermi gas in an axially symmetric
trap slowly rotating around the symmetry axis.     The density $n_s$ of the superfluid
component can be defined as
the rigidity of the superfluid with respect to an infinitesimal  twist in the boundary
conditions\cite{fisher},
$\hat \psi_\sigma({\bf r})\to e^{-i\alpha \phi}
\hat \psi_\sigma({\bf r})$, where   $\phi$ is the azimuthal angle with respect to the symmetry axis,  $\hat\psi_\sigma({\bf r})=\sum_j \hat c_{j\sigma}\varphi_j({\bf r})$, and $\varphi_j({\bf r})$ are the single particle wave functions. The twist generates a term $- \alpha \hat J$ in
the Hamiltonian, where $\hat J$ is the current operator. The resulting supercurrent $J_s=n_s\alpha/m$  can be expressed in terms of correlation functions (\ref{greenfns1},\ref{greenfns2}) using the standard linear response theory. 
We find $n_s=n/2$, where $n$ is the particle density in the normal state.
A similar calculation for
nonzero $\Delta_\infty$ yields $n_s=n$ -- the superfluid density in the gapped steady state is
the same as  in the BCS ground state. Since $n_s$ is the second derivative of the free energy with respect to $\alpha$, the jump   from $n$ to $n/2$ at the
critical point   agrees with the second order character of the
transition.  The reduction in $n_s$ in the  state with zero $\Delta_\infty$ is  consistent with vanishing of the
gap in this state\cite{Schrieffer64}.

We see that the superfluid density is finite in the gapless steady state. This reflects the existence
of pair correlations \re{greenfns2} between fermions, a property that in equilibrium has been directly linked
to nonzero $n_s$ \cite{odlro}. This situation is similar to
  the phenomenon of gapless superconductivity in metals\cite{Parks}, where the superconducting state is also characterized
by non-vanishing anomalous Green's functions and zero spectral gap.
A gapless state
in  superconductors is usually a consequence of strong perturbations - as the perturbation is increased, the
superconducting metal first goes into a gapless state with finite $n_s$ before it becomes
normal.

A crude qualitative understanding of the dynamical transition to the gapless steady state can be derived
from the following thermodynamic argument. After the change of coupling  the initial state \re{initsp} has  energy ${\cal E}(\Delta_f,\Delta_i)>0$ relative to the ground state with gap $\Delta_f$. In thermal equilibrium the system would have the same energy at a certain temperature $T_0(\Delta_f,\Delta_i)$. Let us keep $\Delta_i$ fixed and vary $\Delta_f$. At $\Delta_f=\Delta_i$, $T_0=0$, while the critical temperature $T_c\approx 0.57\Delta_f>0$.  As $\Delta_f$  increases both $T_0$ and $T_c$ grow, but $T_c$ grows faster and $T_0$
never catches up with it. On the other hand, when $\Delta_f$  decreases, $T_c$ also decreases, while $T_0$ grows.
Evaluating $T_0(\Delta_f,\Delta_i)$, to the first order in the coupling constant
we find that $T_0\geq T_c$ for $\Delta_f/\Delta_i\leq 0.52+0.17\lam$ and $T_0< T_c$ otherwise.
Here $\lam=g/d$ is the dimensionless BCS coupling constant  and $d=\langle \eps_{j+1}-\eps_j\rangle$ is the mean level spacing. We see that decreasing the
coupling beyond a certain value  provides enough energy for the transition to the normal state to occur in
thermal equilibrium.

 Let us discuss experimental manifestations of the dynamical transition to the gapless steady state.
 After the initial sweep ($g_i\to{g_f}$) on the BCS side of the resonance such that $\Delta_f/\Delta_i< e^{-\pi/2}$, the gapless state is reached on a
  timescale $\tau_\Delta=1/\Delta_i$.
The pairing gap measured in Ref.~\cite{exp2} is the
energy cost $\Delta E$ of breaking a Cooper pair. This corresponds\cite{Anderson58} to removing  the spin closest to the Fermi level in \re{bcs}. The minimum energy cost is $\Delta E=\min|{\bf b}(\eps)\cdot {\bf s}(\eps)|=\Delta_\infty\cos\theta(0)$.
Using Eq.~\re{sin22}, we obtain $\Delta E=\Delta_\infty \sqrt{1-4\ln^2(\Delta_i/\Delta_f)/\pi^2}$.
  In the steady state, $\Delta E=0$, i.e.  no gap in the rf absorbtion spectrum will be observed.  Molecular condensate  fraction \cite{exp1,exp4} {\it vanishes} in the
gapless state, since $N_{k=0}/N\propto\Delta_\infty^2$ \cite{diener04}.

In conclusion, we determined the dynamics of the paired state of cold atomic fermions following an abrupt lowering of the pairing strength,
$g_i\to g_f$.
On a short $1/\Delta_i$ timescale, where $\Delta_i$ is the equilibrium gap for the initial   coupling
$g_{i}$, the system goes to a non-stationary steady state with a constant gap $\Delta_\infty$ and a distribution function
given by Eqs.~(\ref{lessinf},\ref{sin22}). When the coupling
is reduced so that $\Delta_f/\Delta_i$ is smaller than the critical value of $e^{-\pi/2}$, the steady state gap vanishes. The decay
law of the time-dependent order parameter $\Delta(t)$ changes from power law to exponential as $\Delta_f/\Delta_i$ goes through
the critical point, Eqs.~(\ref{asymp},\ref{asymsm}). The gapless  state combines features of normal  and superfluid states. In particular, the
gap and the condensate fraction vanish, while the superfluid density is nonzero.

We thank E. Abrahams, B. Altshuler, and J. Schmalian for useful discussions. M. D. was in part supported by
the U.S. Dept. of Energy, Office of Science, under Contract
No.W-7405-ENG-82 and by the ICAM.
After this work was completed, we became aware of a work by  Barankov and  Levitov\cite{BL2006} that
reports  some results similar to ours.

\end{document}